\documentclass[11pt]{article}
\usepackage{graphicx}
\usepackage{epsfig}

\begin{document}

\title{\bf {Form Factors in Relativistic Quantum Mechanics Approaches and 
Space-Time Translation Invariance}}

\author{ B.  DESPLANQUES\thanks{{\it E-mail address:}  
desplanq@lpsc.in2p3.fr}  \\  
Laboratoire de Physique Subatomique et de Cosmologie \\
(UMR CNRS/IN2P3 -- UJF -- INPG), 
  F-38026 Grenoble Cedex, France }

\maketitle

\begin{abstract}
\small{Invariance of form factors under Lorentz boosts is a criterion often advocated
to determine whether their estimate in a RQM framework is reliable. 
It is shown that verifying  relations stemming from covariance properties 
under space-time translations could be a more important criterion. 
Form factors calculated in different approaches for a simple system are 
discussed with these various respects. An approximate method is shown to remove 
the main discrepancies related to a violation of the above relations.}
\end{abstract}
%%%%%%%%%%%%%%%%%%%%%%%%%%%%%%%%%%%%%%%%%%%%%%%%%%%%%%%%%%%%%%%%%%%%%%%%%%%%%%%%
\section{Introduction}
There are many ways to implement relativity in the description of a few-body
system and its properties, such as form factors. Ultimately, their predictions 
should converge but in an approximate calculation, some approaches may be 
more efficient than other ones. Relativistic quantum mechanics (RQM) has 
the advantage over field theory that it is dealing with a fixed number 
of (effective) degrees of freedom. As a counterpart, relativistic covariance 
properties are not trivially fulfilled. 
Following Dirac's work\cite{Dirac:1949cp}, several approaches depending 
on the symmetry properties of the hyper surface on which physics is formulated
have been proposed: they are the instant, front, and point forms. 
These ones have been used for calculating form factors of various systems, 
evidencing large discrepancies in the approximation of a one-body 
current (see Ref. \cite{Amghar:2002jx} and references therein). 
This raises the question of determining criteria allowing one to discriminate
between the various approaches. Invariance under Lorentz boosts, which can 
be easily checked, is obviously one of them but relativity also implies other
transformations such as space-time translations. The invariance under 
these ones provides the well-known conservation of energy and momentum, 
which, evidently, holds globally for the system. However, in an incomplete 
RQM calculation, this property is not necessarily fulfilled at the interaction 
vertex of its constituents with an external probe. 

In the present contribution, we examined form factors of a simple system with
respect to the above transformations. After showing the results obtained in
different approaches (Sec. 2), we proceed to their discussion in view 
of relations which stem from the transformation of currents under space-time
translations (Sec. 3).
%%%%%%%%%%%%%%%%%%%%%%%%%%%%%%%%%%%%%%%%%%%%%%%%%%%%%%%%%%%%%%%%%%%%%%%%%%%%%%%%
\section{Form Factors in Different Forms: Single-Particle Approximation}
The system whose form factors are studied here is a theoretical one. It
corresponds to the ground state of the Wick-Cutkoski model (scalar particles
exchanging a massless scalar meson). Due to a hidden symmetry, the 
Bethe-Salpeter equation for this system can be easily solved and the 
charge and scalar form factor can be calculated exactly. In some sense, 
this provides our ``experiment". It can be compared to form factors 
calculated in the single-particle current approximation for different 
RQM approaches and using the same solution of a mass operator.
The two types of contributions are shown in Fig. 1, l.h.s. and r.h.s.
respectively. 
\begin{figure}[htb] 
\hspace*{5mm}\epsfig{file=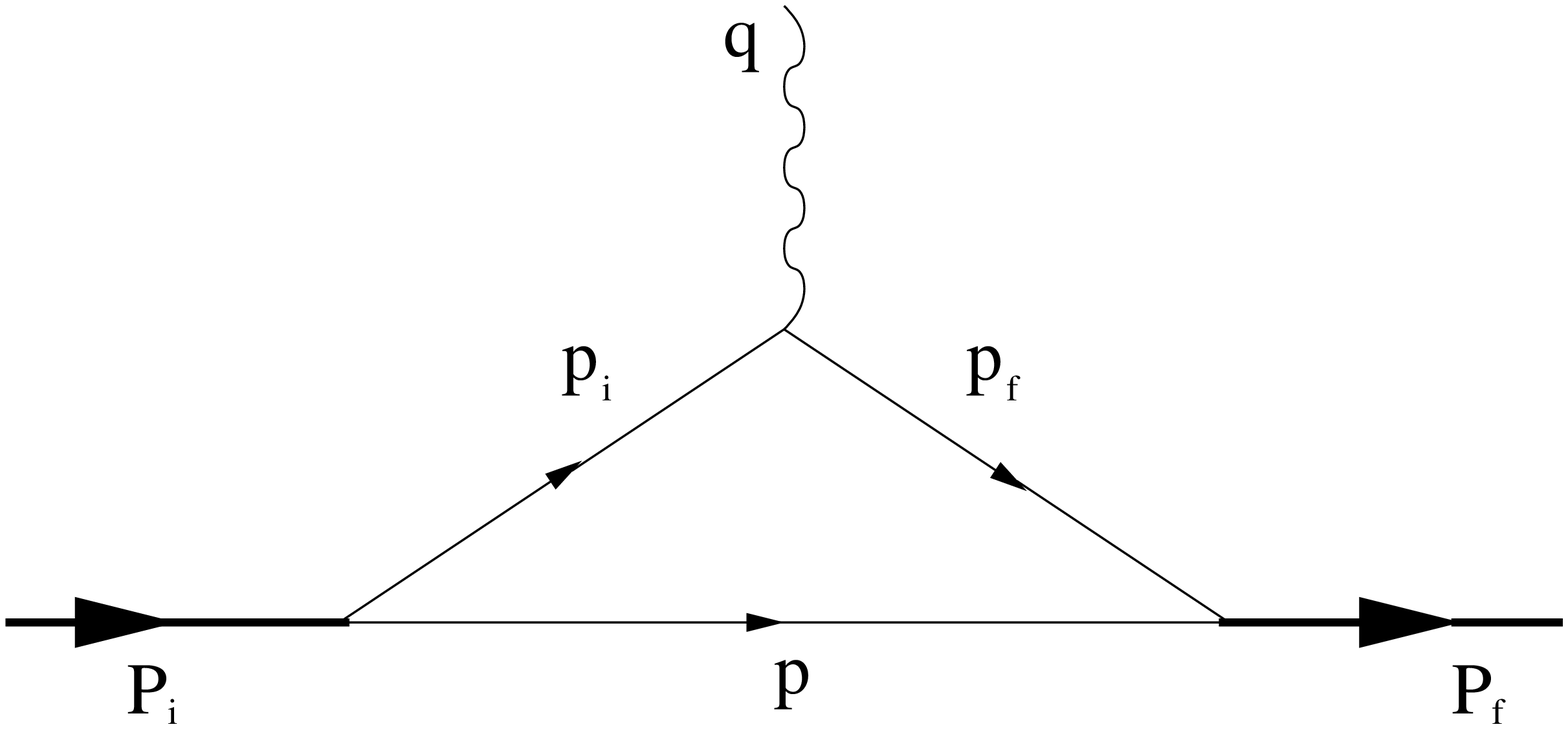, width = 7.0cm} \hspace*{5mm}
\epsfig{file=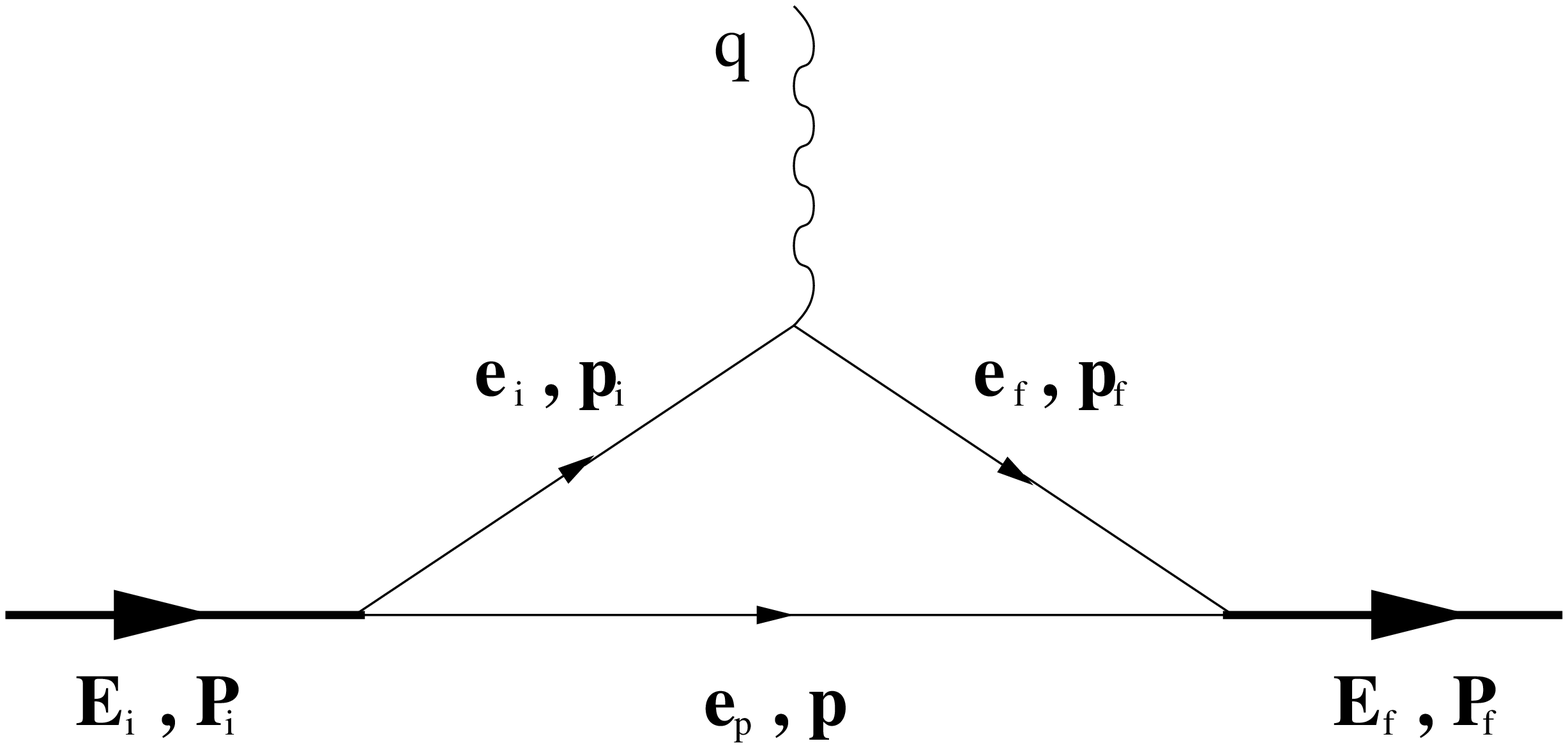, width = 7.0cm}   
\caption{Single-particle current contribution in field-theory  (l.h.s.) 
and RQM (r.h.s.) approaches. Intermediate particles are respectively 
off-mass shell and on-mass shell.}
\end{figure}

In the instant form, where results are not invariant under a Lorentz 
transformation (boost), form factors can be considered for the Breit 
frame (I.F. (Breit frame)) and for a frame where the initial- and final-state 
momenta are parallel while their sum goes to infinity (I.F. (parallel)). 
In the front form, where results are not invariant under a Lorentz 
transformation (rotation), they can be considered for the configuration $q^+=0$ 
(F.F. (perp.)) and for a configuration where the initial- and final-state 
momenta are parallel to the front orientation (F.F. (parallel)). Not
surprisingly, the last results are identical to those in the instant form 
with an infinite average momentum. In the point-form, 
results turn out to be Lorentz invariant. Different versions may be
nevertheless considered. The first one\cite{Bakamjian:1961,Sokolov:1985jv}, 
extensively used in many works, is an instant form displaying 
the same symmetry as the point form (``P.F."). 
The second one\cite{Desplanques:2004rd} is more in the spirit 
of the Dirac point form, 
where physics is formulated on an hyperboloid surface (D.P.F.).
 
Results are presented in Fig. 2 for the charge form factor, $F_1(Q^2)$,
at both low and high $Q^2$. These ranges  are aimed to point to 
the charge radius, $ <r^2>$, and to the asymptotic behavior 
(expected to be $Q^{-4}$).
\begin{figure}[htb]
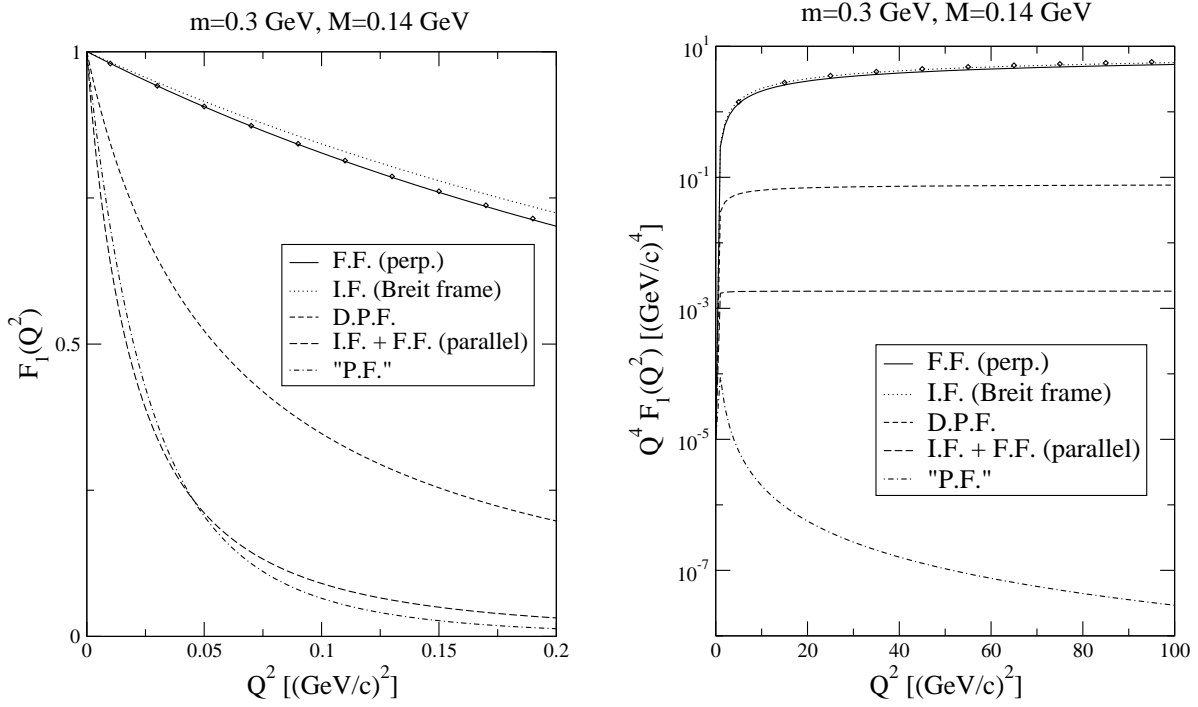

\epsfig{file=fram1s.eps, width = 7.5cm} \hspace*{5mm}
\epsfig{file=fram1x.eps, width = 7.5cm}   
\caption{Form factors at low and large $Q^2$  (l.h.s. and r.h.s.
respectively). Parameters are appropriate to a pion-like system. See text for
the definition of abbreviations.}
\end{figure} 
In comparison to ``experiment" (small diamonds in the figure), it is noticed that
the instant-form (Breit-frame) and  front-form ($q^+=0$) cases 
do rather well. Lorentz invariance, which underlies point-form 
results, does not imply good results. Apart from the above instant- 
and front-form cases, the charge radius scales like the inverse 
of the mass of the system, leading to the paradox that the former quantity  
goes to infinity when the latter goes to zero. 
This suggests the violation of some symmetry. Which one however?
%%%%%%%%%%%%%%%%%%%%%%%%%%%%%%%%%%%%%%%%%%%%%%%%%%%%%%%%%%%%%%%%%%%%%%%%%%%%%%%%
\section{Role of Space-Time Translation Invariance }
Space-time translation invariance implies the 4-momentum conservation. 
This has to hold globally for any physical process, giving the relation 
($P_i^{\mu}+q^{\mu}=P_f^{\mu}$ in the present case (Fig. 1)). In field-theory, 
such a relation also holds at the interaction vertex of the constituents 
with the external probe ($p_i^{\mu}+q^{\mu}=p_f^{\mu}$). However, the last one 
is not verified in RQM approaches, where constituents are on-mass shell.
To quantitatively discuss the consequences of this feature, it is appropriate 
to consider relations that stem from the transformation of a current under
space-time translations\cite{Lev:1993}, which are generated by
the operators of the Poincar\'e algebra, $P^{\mu}$. Considering matrix elements, 
one should verify  relations like:
\begin{eqnarray}
\langle \;|\Big[ P^{\mu}\;,\; J^{\nu}(x)\Big]|\; \rangle =
-i\langle\; | \partial^{\mu}\,J^{\nu}(x) |\;\rangle,\hspace*{6mm}
\nonumber \\ 
\langle\;| \Big[P_{\mu}\;,\Big[ P^{\mu}\;,\;  J^{\nu}(x)\Big]
\Big] |\;\rangle = 
-\langle \;| \partial_{\mu} \partial^{\mu}\,J^{\nu}(x) |\;\rangle \, , 
\nonumber  \\  {\rm or,\;here,} \;\;\;
\langle\;| q^2\; J^{\nu}(x)| \;\rangle =
\langle\;| (p_i-p_f)^2\; J^{\nu}(x)|\; \rangle\,, \hspace*{4mm} 
\end{eqnarray}
which are automatically fulfilled in field-theory but suppose 
the existence of many-body currents in RQM approaches. 
Assuming a single-particle current, it can be verified that the 
last equation is fulfilled exactly in the front-form ($q^+=0$) case or
approximately in the instant-form (Breit frame) one. 
In all other cases, it is violated by large factors, from 30 up to 35000
 (``P.F.") here.

\begin{figure}[htb]
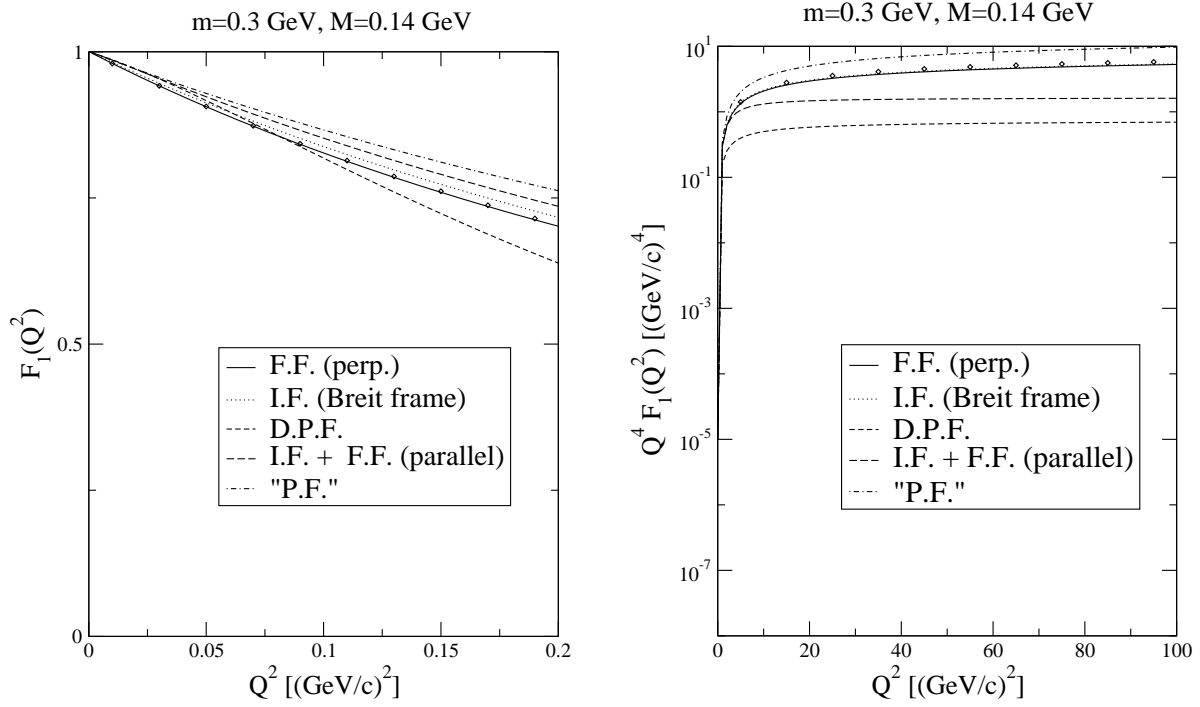

\epsfig{file=fram1sr.eps, width = 7.5cm} \hspace*{5mm}
\epsfig{file=fram1xr.eps, width = 7.5cm}   
\caption{Form factors with corrections aimed to fulfill relations stemming 
from transformations of currents under space-time translations, at low 
and large $Q^2$  (l.h.s. and r.h.s. respectively).}
\end{figure} 

Examination of these cases indicates that the factor multiplying $Q^2$ in
calculations misses interaction effects. Correcting this factor  
such as to remove the above violation factors therefore provides 
an approximate way to account for the missing many-body currents 
required at all orders in the interaction to fulfill Eqs. 1.
This approach has been applied to the form factors shown in Fig. 2. 
The resulting ones are presented in Fig. 3. It is observed that the largest 
discrepancies have vanished at low and high $Q^2$. In particular, the peculiar
behavior of form factors in the limit of a zero-mass system is completely
removed.
%%%%%%%%%%%%%%%%%%%%%%%%%%%%%%%%%%%%%%%%%%%%%%%%%%%%%%%%%%%%%%%%%%%%%%%%%%%%%%%%
\section{Conclusion}
The present work shows that constraints from  space-time translation  
transformations could be more important than those from (homogeneous) 
Lorentz transformations. Invariance under these last ones turns out 
to be a disadvantage as there is no frame where the violation of the other
constraints can be minimized. To a large extent, the present work confirms
the standard front-form approach ($q^+=0$) and instant-form one 
(Breit-frame or $E_i=E_f$ case more generally) as more reliable frameworks
in the single-particle current approximation.

\end{document}